\documentclass[useAMS,usenatbib]{mn2e}

\usepackage{graphics}
\usepackage{rotating}
\usepackage{multirow}

\title[DETECTION PROBABILITY OF A LOW-MASS PLANET FOR
TRIPLE LENS EVENTS]{DETECTION PROBABILITY OF A LOW-MASS PLANET FOR
TRIPLE LENS EVENTS: IMPLICATION OF PROPERTIES OF BINARY-LENS
SUPERPOSITION}

\author[Ryu et al.]{Yoon-Hyun Ryu$^1$\thanks{E-mail:
yhryu@astroph.chungbuk.ac.kr (YHR); hyc@knu.ac.kr (HYC);
mgp@knu.ac.kr (MGP)}, Heon-Young
Chang$^2$\footnotemark[1]\thanks{corresponding author}, and Myeong-Gu Park$^2$\footnotemark[1]\\
$^{1}$Department of Physics, Chungbuk
National University, Cheongju 361-763, Korea\\
$^{2}$Department of Astronomy and Atmospheric Sciences, Kyungpook
National University, Daegu 702-701, Korea}
\begin{document}


\maketitle

\label{firstpage}
\begin{abstract}
In view of the assumption that any planetary system is likely to be
composed of more than one planet, and the multiple planet system
with a large mass planet has more chance of detailed follow-up
observations, the multiple planet system may be an efficient way to
search for sub-Jovian planets. We study the central region of the
magnification pattern for the triple lens system composed of a star,
a Jovian mass planet, and a low-mass planet to answer the question
if the low-mass planet can be detected in high magnification events.
We compare the magnification pattern of the triple lens system with
that of a best-fitted binary system composed of a star and a Jovian
mass planet, and check the probability in detecting the low-mass
secondary planet whose signature will be superposed on that of the
primary Jovian mass planet. Detection probabilities of the low-mass
planet in the triple lens system are quite similar to the
probability in detecting such a low-mass planet in a binary system
with a star and only a low-mass planet, which shows that the
signature of a low-mass planet can be effectively detected even when
it is concurrent with the signature of the more massive planet,
implying that the binary superposition approximation works over a
relatively broad range of planet mass ratio and separations, and the
inaccuracies thereof do not significantly affect the detection
probability of the lower mass secondary planet. Since the signature
of the Jovian mass planet will be larger and lasts longer, thereby
warranting more intensive follow-up observations, the actual
detection rate of the low-mass planet in a triple system with a
Jovian mass can be significantly higher than that in a binary system
with a low-mass planet only. We conclude that it may be
worthwhile to develop an efficient algorithm to search for
`super-Earth' planets in the paradigm of the triple lens model for
high-magnification microlensing events.
\end{abstract}


\section{INTRODUCTION}

To detect and characterize extrasolar planets various techniques
have been employed, including the radial velocity technique
\citep{may95}, the transit method \citep{cha07}, direct imaging
\citep{cha04}, pulsar timing analysis \citep{wol92}, and
microlensing
\citep{bon04,uda05,bea06,gou06,ben08,gau08,don09,jan10,sum10}.
Compared to other techniques, the microlensing method has an
important advantage of being applicable to planets to which other
methods are generally insensitive; the microlensing technique is
sensitive to detecting low-mass planets and cool planets, or even
free-floating planets \citep{ben02,han04,han05b}. This sensitivity
is so important for testing the core accretion theory of planet
formation which predicts that the dominant planets in any planetary
system should form in the vicinity of the `snow line', which is
located at a few AU from the host star \citep{lau04,ida05,ken06}.

When a microlensing event occurs by a planetary system, the
planetary signal appears as a short-duration perturbation to the
standard light curve induced by the lens star \citep{mao91,gou92,ben96}.
The planetary lensing signal induced by a planet with a mass of
Jupiter lasts for a duration of $\sim 1$ day and of $\sim 1.5 $
hours for a mass of Earth. Therefore, the discovery of a terrestrial
planet would only be possible by high-cadence anomaly monitoring. In
fact, the detection of a significant number of terrestrial extra-solar
planets requires well-coordinated efforts involving a network (e.g.,
MOA-II: Sumi et al. 2010; OGLE-IV: Udalski et al. 2005). 
MOA-II has reported two low-mass planets with their survey,
MOA-2007-BLG-192Lb (Bennett et al. 2008) and
OGLE-2007-BLG-368Lb (Sumi et al. 2010), and is preparing
one, MOA-2009-BLG-266Lb (in preparation). 
The planetary perturbation occurs when the source star
crosses the caustic or passes close to it. Caustic-crossing events
cause conspicuous double-peaks over the smooth light curve induced
by a lensing star. However, the perturbations due to
caustic-crossing occur without any prior warning so that current
microlensing follow-up observations are focusing on high
magnification events for the sake of practicality. For
high-magnification events, the source trajectories always pass close
to the perturbation region around the central caustic induced by the
planet and thus timing can be predicted fairly accurately
\citep{gri98,han01b,bon02,rat02,yoo04}. \citet{gri98} have shown
that planets with masses as low as $10 M_{\earth}$ could be detected
with significant probability in events with magnifications $\sim 50$
by monitoring the peaks of the events with a photometric precision
of $\sim 1$ per cent. Rhie et al.(2000) have first showed that high
magnification events are sensitive to low-mass (Earth-mass)
planets (For the lower limit of the planet mass most recently
reported, see Yee et al. 2009).

Since the discovery of the first extra solar planet orbiting a
solar-type star \citep{may95}, the Extra-solar Planet
Encyclopedia\footnote{{\rm http://exoplanet.eu}} lists 429 entries,
including 45 multiple planetary systems, as of the 09th of February
in 2010. The detectable mass of exo-planets is extending below the
$10 M_{\earth}$ regime, with the discoveries of Gliese 876d at a
mass of $\sim 7.5 M_{\earth}$ \citep{riv05}, three planets around HD
40307 with masses of $\sim 4.2$, $\sim 6.9$ and $\sim 9.2
M_{\earth}$ \citep{may09a}, and Gliese 581e with a mass of $\sim 1.9
M_{\earth}$ \citep{may09b} using the radial-velocity technique, as
well as OGLE 2005-BLG-390Lb detected by microlensing at a mass of
$\sim 5.5 M_{\earth}$ \citep{bea06,ben08}. Many of the discovered
`super-Earth' planets have been revealed through the close
re-examination of planetary signals that already proved the
existence of their big brother. Considering that a planetary system
is likely to be composed of more than one planet, this kind of
strategy to find low-mass planets in the multiple planet systems may
become an efficient way to search for terrestrial planets in the
sense that it is easier to detect subtle signals when one knows
where to look for.

The magnification pattern due to the triple-lens systems is known to
be well approximated by the superposition of the magnifications due
to individual planets for planetary caustics (Han et al. 2001) and
for central caustics (Han 2005). Therefore, we may expect that the
detection probability of low-mass planets in a triple system with a
Jovian mass planet plus a low-mass planet should be quite similar to
the detection probability in a binary system with only a low-mass
planet. However, this is known only for a limited range of planet
mass ratio and separations (Han 2005), and also the effect of the
deviation from the superposition approximation to the detection
probability is not known. Moreover, it is not clear if the smaller
deviation in the magnification due to a low-mass planet superposed
on the larger deviation due to a more massive Jovian mass planet can
be effectively detected in high magnification events when the
`adjusted' binary lens model is fitted to incorporate the additional
deviation from the low-mass planet, possibly removing the signature
of the low-mass planet. So in this paper, we study the triple lens
system (a lens star, a Jovian mass planet, and a low-mass planet)
for a broader range of planetary masses and separations than in
previous studies, motivated by the fact that all planets in the
lensing zone will substantially affect the central caustics and thus
the existence of the multiple planets can be inferred by analyzing
additionally deformed anomalies in the light curve of high
magnification microlensing events \citep{gau98,gau08}. We further
calculate the probability to detect the low-mass planet in the
high-magnification events of a triple lens system and compare that
to that of a binary lens system (a lens star and a low-mass planet).
Since the dependence of the size of the central caustic on the
planet/star mass ratio $q$ is linear \citep{gri98,chu05,han06}, detecting
signals of Earth-mass planets with $q \sim 10^{-5}$ from the
planet-search strategy of monitoring high-magnification event of the
binary lens system is observationally challenging
\citep[e.g.,][]{ben96,ben02,gau00}. Therefore, it should be crucial
and timely to address the question if the detection probability of a
low-mass planet for high-magnification events of the triple lens
system is comparable to that of the binary lens system. Answers to
this question may well have implications on interpretations of the
light curve of high-magnification microlensing events. Indeed,
recently there have been efforts to re-analyze the observational
data along the line of this point. For example, \citet{kub08} have
attempted to estimate the detection probability for the secondary
planet by re-analyzing the observational data of OGLE 2005-BLG-390.
Their assumptions may be justified for their purposes in that the
impact parameter of OGLE 2005-BLG-390 is large, i.e., $u_0=0.359$.
On the contrary, in this paper we specifically concentrate on the
high magnification microlensing events ($ u_0\lid0.01$).

In \S 2, we begin with a brief description of the multiple lens
systems. In \S 3, we construct the fractional deviation of the
magnification map induced by the secondary planet. In \S 4, we
present the detection probability and compare the triple lens case
with the binary lens case. In \S 5, we conclude with a summary of
our results.

\section{MULTIPLE LENS SYSTEMS}
When a source star is gravitationally lensed by a point-mass lens,
the source is split into two images with the total magnification
given as a simple analytic form of
\begin{equation}
A_0 = {u^2+2\over u\sqrt{u^2+4}},
\end{equation}
where $u$ is the lens-source separation normalized by the angular
Einstein ring radius \citep{pac86}.  The angular Einstein ring
radius is related to the physical parameters of the lens by
\begin{equation}
\theta_{\rm E} = \sqrt{4GM \over c^2}
       \left( {1\over D_{\rm ol}} - {1\over D_{\rm os}} \right)^{1/2},
\end{equation}
where $M$ is the mass of the lens and $D_{\rm ol}$ and $D_{\rm os}$
represent the distances from the observer to the lens and source,
respectively.

If the lens star accommodates a planetary companion, the latter may
further perturb the image and so changes the magnification. When a
lensing event is caused by the multiple lens system, locations of
the individual image are obtained by solving the lens equation
expressed in complex notations by
\begin{equation}
\zeta=z+\sum_{j=0}^{N-1} \frac{q_j}{\bar{z}_j-\bar{z}},
\end{equation}
where $q_j$'s and $z_j$'s represent the mass fractions of individual
lenses ($j=0$ for the central lens star, $j=1$ for the primary
planet, and $j=2$ for the secondary planet and so on) such that
$\sum_j q_j =1$ and the positions of the lenses, respectively,
$\zeta=\xi+i\eta$ and $z=x+iy$ are the positions of the source and
images, and $\bar{z}$ denotes the complex conjugate of $z$
\citep{wit90}.  Note that all these lengths are normalized by the
combined Einstein ring radius, which is equivalent to the Einstein
ring radius of the single lens with a mass equal to the total mass
of the system. The total magnification is the sum of magnifications
of the individual images, $A=\Sigma_k A_k$. The magnification of
each image evaluated at the position of each image $z_k$, is given
by
\begin{equation}
A_k=\frac{1}{|{\rm det}\ J|_{z_k}},
\end{equation}
where
\begin{equation}
    {\rm det}\ J=1-{\partial \zeta \over \partial \bar{z}}
     {\overline{\partial \zeta} \over \partial \bar{z}}.
\end{equation}

The fundamental difference in the geometry of the multiple lens
system from that of the single point-mass lens system is the
formation of caustics.  The caustic refers to the source position on
which the magnification of a point source event becomes infinity,
i.e., det $J=0$. The set of caustics forms closed curves. For
example, for the planetary lensing system composed of a single star
and a single planet, there exist two sets of disconnected caustics:
One `central' caustic is located close to the lens star, while the
other `planetary' caustic(s) is (are) located away from the lens
star. The number of the planetary caustics is one or two depending
on whether the planet lies outside or inside the Einstein ring. When
the planetary companion is close to the Einstein ring, the planetary
and central caustics merge into a single `resonant caustic'. The
central caustic plays a crucial role in current microlensing planet
searches, as stated earlier, in that the central caustic due to the
planet has a high probability of
perturbing the light curve of high-magnification event,
therefore detectable as the planetary signal. For the triple lens
case, the lens equation becomes a rather complicated tenth order
polynomial equation that has 4, 6, 8, or 10 solutions, corresponding
to physical images depending on the configuration of the lens system
and the location of the source (Rhie 2002).
In the triple lens system, patterns
of the planetary caustics may well be described by the superposition
of those of the single-planet systems \citep{han01a,han05a}. They
are barely affected by each other unless the projected positions of
the planets are close. On the contrary, in the central region of the
magnification map the anomaly pattern induced by one planet can be
significantly affected by the existence of another planet.

\section{DEVIATIONS DUE TO SECONDARY PLANET}

In Fig.~2, we show the fractional deviation of the magnification map
obtained by fitting the binary lens model to the magnification map
induced by triple lens systems, which is defined by
\begin{equation}
\varepsilon\equiv\frac{A_{\rm tri}-A_{\rm bin}}{A_{\rm bin}},
\end{equation}
where $A_{\rm tri}$ and $A_{\rm bin}$ represent the magnification
map generated with the primary plus secondary planets and that
obtained by fitting the binary lens model in which the secondary
planet is absent, respectively. The inverse ray-shooting technique
is used to obtain the magnification map \citep{sch86,kay86,wam97}.
Deviation maps are presented in terms of the lens position of the
secondary planet both in angles between position vectors of two
planets $\theta$ and in separations $s_2$ (see Fig.~1). Each map of
the fractional deviation is calculated as a function of the source
position $(\xi,\eta)$. The coordinates are set so that the lens star
is at the center. We concentrate only on regions of $|\xi| \lid
0.01$ and $|\eta| \lid 0.01$ for which high-magnification events are
relevant. In generating the magnification map induced by triple lens
systems the primary planet is fixed to be located on the $\xi$ axis,
$(s_{1x},s_{1y})=(-1.3,0.0)$. We set the mass ratio of the primary
planet to the lens star $q_1=3\times10^{-3}$ and that of the
secondary planet $q_2=1\times10^{-3}$, corresponding to a
Jupiter-mass and Saturn-mass planets orbiting a $0.3 M_{\odot}$
star, respectively. Unless otherwise stated, in all computations
throughout this paper we adopt typical values of distances for
Galactic bulge events:  $D_{\rm ol}=6$ kpc, $D_{\rm os}=8$ kpc. We
also assume the source radius $\theta_\star$ to be
$\theta_\star/\theta_E$ = 0.001, which corresponds to a typical
main-sequence star. Note that in the fitting procedure parameters of
the binary lens model composed of the a lens star and the primary
planet are constrained to find their best fit values. To minimize
the difference between the epsilon we generated using the triple
lens and one we modeled by the binary lens model, the minimization
process has been performed by the least square method with varying
the mass ratio of the primary planet, $q_1$, and its position,
$s_1$. Contours are drawn at the levels of $\varepsilon=\pm 1\%$,
$\pm5\%$, $\pm10\%$, and $\pm20\%$, and the regions of negative
$\varepsilon$ are shown in blue and positive $\varepsilon$ in red.
The color changes into light shade as $|\varepsilon|$ level
decreases.

As one may expect, distorted regions due to the secondary planet are
confined to the central caustic of the primary planet due to the
nonlinear interference between perturbations produced by two
planets. On the other hand, it should be noted that when
$\theta=0\degr$ there may exist a degenerate case where two-planet
geometries will give rise to exactly the same magnification map with
a single planet of a slightly larger mass as previously noted by
\citet{gau98}. In our particular example it can be seen in the case
of $s_2=1.26$ since the primary planet is located at (-1.3, 0.0). We
also note that the degeneracy between $s_2$ and $s^{-1}_2$ can be
seen due to the additional planet. Regions deformed by the secondary
planet revolve as the angle between two planets varies, and become
broader as  $s_2$ approaches unity. To explore the effect of $q_2$,
we show deviation maps for several $s_2$ and $q_2$ in Fig.~3. We set
the mass ratio of the primary planet to the lens star
$q_1=3\times10^{-3}$ and the angle between position vectors of two
planets $\theta=30\degr$. As in Fig.~2, contours are drawn at the
levels of $\varepsilon=\pm 1\%$, $\pm5\%$, $\pm10\%$, and $\pm20\%$,
and the regions of negative $\varepsilon$ are shown in blue and
positive $\varepsilon$ in red. From the maps, one finds that the
lower limit of the mass of the secondary planet that may cause
sufficient deformations in the magnification map is between
$1\times10^{-5} \la q_2 \la 1\times10^{-4}$, in which
$q_2=1\times10^{-4}$ corresponds to a $10 M_{\earth}$ planet
orbiting a $0.3 M_{\odot}$ star. In other words, an Earth-mass
planet can only be possibly detectable in a narrow region of the
parameter space unless the photometric accuracy allows one to
examine the observational data in the levels of $|\varepsilon| \sim
1\%$.

\section{COMPARISON OF DETECTION PROBABILITY}

To `discover' a planet, one may demand that $\Delta \chi^2$ between
the light curve calculated by the planetary  lens model and the
observed data should be smaller than a carefully chosen critical
value. Or, as another way to disclose a planet the $\Delta
\chi^2$-based criterion can be translated into so-called Gould \&
Loeb criterion, assuming a predetermined photometric accuracy and
non-white noise \citep{gou92}. Gould \& Loeb criterion considers
deviations as a planetary signal when a few observational points are
consecutively deviated from the single-lens light curve. In the
current exercise, we are going to follow Gould \& Loeb criterion.
Thus, we will count a signal of planetary detection as successful if
at least one point in the deviation map has the deviation amplitude
larger than a certain threshold. We have defined the detection
probability of the low-mass planet as follows: Firstly, for given
$q_2$, $s_2$ and $\theta$, we calculate the area in which the value
of $|\varepsilon|$ in the deviation map, e.g., shown in Fig.~2 and
3, is greater than 5\%. Bearing in mind that monitoring high-magnification
events should be an efficient strategy of detecting `super-Earth'
planets in view of practical purposes, 
we only consider this quantity in the range of $|u_0| \lid 0.01$ as
commonly employed.
Having done that, we average the obtained area over the
position angle in the range of $0\degr\lid\theta\lid180\degr$. Then,
we normalize the averaged area with the area of $|u_0| \lid 0.01$.
In this way, we obtain the detection probability as a function of
$q_2$ and $s_2$. Note that we repeat the same calculation for the
binary lens with a planet that has the same mass as the secondary
planet in the triple lens system, except that we construct
fractional deviation maps given as
\begin{equation}
\varepsilon\equiv\frac{A_{\rm bin}-A_{\rm single}}{A_{\rm single}},
\end{equation}
where $A_{\rm bin}$ and $A_{\rm single}$ represent the magnification
map generated with the binary lens system having a planet and that
obtained by fitting of the single lens model, respectively.

In Fig.~4, we compare the probability of detecting the low-mass
planet in triple systems with that in binary systems, provided that
the detection threshold is $|\varepsilon| > 5\%$. The detection
probability of the secondary low-mass planet in the triple lens
system is presented by grey-scale  such that darker shade represents
higher probability as shown in the scale-bar. We note that for a
given $q_2$ the probability in detecting the secondary planet
becomes higher as $s_2$ approaches unity, as expected in plots such
as Fig.~2. We find that in the triple lens system the detection
probabilities of the low-mass planet are $\sim 50\%$, $\sim 10\%$,
and $\sim 1\%$ for $q_2= 10^{-3}, 10^{-4}$, and $10^{-5}$,
respectively, if we only consider the secondary planet residing in
the lensing zone, $0.6 \la s_2 \la 1.6$. For comparison, we also
present as dotted contours the probability of detecting the low-mass
planet in binary systems. The detection probabilities for low-mass
planets have been calculated (Gould \& Loeb 1992; Bennett \& Rhie 1996;
Griest \& Safizadeh 1998; Gaudi, Naber, \& Sackett 1998; Rhie et al. 2000;
Bennett \& Rhie 2002; Rattenbury et al. 2002; Kubas et al. 2008).
Interestingly enough,
for a given $q_2$ and $s_2$ the detection probability of the
low-mass planet in the triple lens system is very similar to that in
the binary lens system. It makes sense when recalling that the
magnification pattern due to the triple-lens systems is well
approximated by the superposition of the magnifications due to the
individual planets (Han et al. 2001; Han 2005). The fundamental
reason why the detection probability of low-mass planets in a
triple-lens system with a Jovian mass planet could be similar to the
detection probability in a binary-lens system is that in most cases
the binary superposition works well even in the central region. If
the interference between caustics due to two planets destroys its
original shapes then there is no reason that we end up with
residuals implying a second planet. Having said so, another
necessary condition that the detection probability of low-mass
planets in a triple-lens system with a Jovian mass planet could be
similar to the detection probability in a binary-lens system is that
the minimization should finds a solution for the Jovian planet
accurately enough.

In Fig.~5, we present the detection probability for the lower
threshold of $|\varepsilon| > 1\%$, provided that photometric
uncertainties of $\gid 1\%$ are achievable in the future. It is
natural to find that the probability in the traditional lensing zone
becomes higher. That is, we find that in the triple lens system the
detection probabilities of the low-mass planet  are $\sim 80\%$,
$\sim 40\%$, and $\sim 5\%$ for $q_2= 10^{-3}, 10^{-4}$, and
$10^{-5}$, respectively, if we only consider the secondary planet
residing in the lensing zone. It is also found that for high-mass
planet of $q_2 \ga 10^{-4}$ corresponding to a $10 M_{\earth}$
planet orbiting a $0.3M_{\odot}$ star there exist reasonably high
probabilities in the broader range than the lensing zone, i.e., $s_2
\la 0.6$ or $s_2 \ga 1.6$. The detection probability of the low-mass
secondary planet in the triple lens system is also comparable to
that in the binary lens system, except when $q_2 \la 3 \times
10^{-6}$ corresponding to a $0.3 M_{\earth}$ planet orbiting a $0.3
M_{\odot}$ star. In that parameter space the detection probability
in the binary lens system is somewhat higher than that in the triple
lens system.

\section{CONCLUSION}

We have computed the detection probability of the low-mass planet
specifically for high-magnification events of the triple lens
system, motivated by the fact that the central caustic is distorted
by any companion to the lens star. Having done that, we have
compared the detection probability of the low-mass secondary planet
for high-magnification events of the triple lens system with that of
the same low-mass planet but in the binary lens system. It should be
stressed that the detection probability of the low-mass planet
for high-magnification events of the triple lens system given in
this paper is a kind of relative probability in the sense that our
criteria on the discovery of a planet is actually closer to one of
signals of the planet. The detection of signals of a low-mass planet
may not be sufficient for the discovery of such a planet. Having
compared results of two cases with the same criteria, however,
probabilities given here are still worthy in lending emphasis to the
light curves of the triple lens system.

Our main findings are as follows:

(1) In the triple lens case where the secondary planet resides in
the lensing zone, the detection probabilities of the low-mass planet
for the high magnification events are $\sim 50\%$, $\sim 10\%$, and
$\sim 1\%$ for $q_2= 10^{-3}, 10^{-4}$, and $10^{-5}$, respectively,
when the detection criterion is $|\varepsilon| >$ 5\% where
$\varepsilon$ is the deviation in magnification.

(2) When the detection criterion is $|\varepsilon|> 1\%$, those
probabilities increase to $\sim 80\%$, $\sim 40\%$, and $\sim 5\%$
for $q_2= 10^{-3}, 10^{-4}$, and $10^{-5}$, respectively. For
high-mass planet of $q_2 \ga 10^{-4}$, there exist reasonably high
probabilities outside the usual lensing zone, $s_2 \la 0.6$ or $s_2
\ga 1.6$.

(3) For a given $q_2$ and $s_2$, the detection probability of the
low-mass planet in the triple lens system is comparable to that in
the binary lens system. Therefore, it is
quite necessary to develop an efficient algorithm search for
`super-Earth' planets in the paradigm of the triple lens model as
well as of the binary lens model.

\section*{Acknowledgments}

We thank the anonymous referees for critical comments which clarify
and improve the original version of the manuscript. We are also
grateful to Ki-Won Lee for many insightful comments and careful
reading of the manuscript. This work is the result of research
activities at the Astrophysical Research Center for the Structure
and Evolution of the Cosmos (ARCSEC) supported by the Korea Science
\& Engineering Foundation (KOSEF). HYC was supported by Basic
Science Research Program through the National Research Foundation of
Korea(NRF) funded by the Ministry of Education, Science and
Technology (2009-0071263).


\clearpage

\begin{figure*}
\begin{minipage}[hpt]{17cm}
\begin{center}
\includegraphics[scale=1.0,angle=0,clip=true]{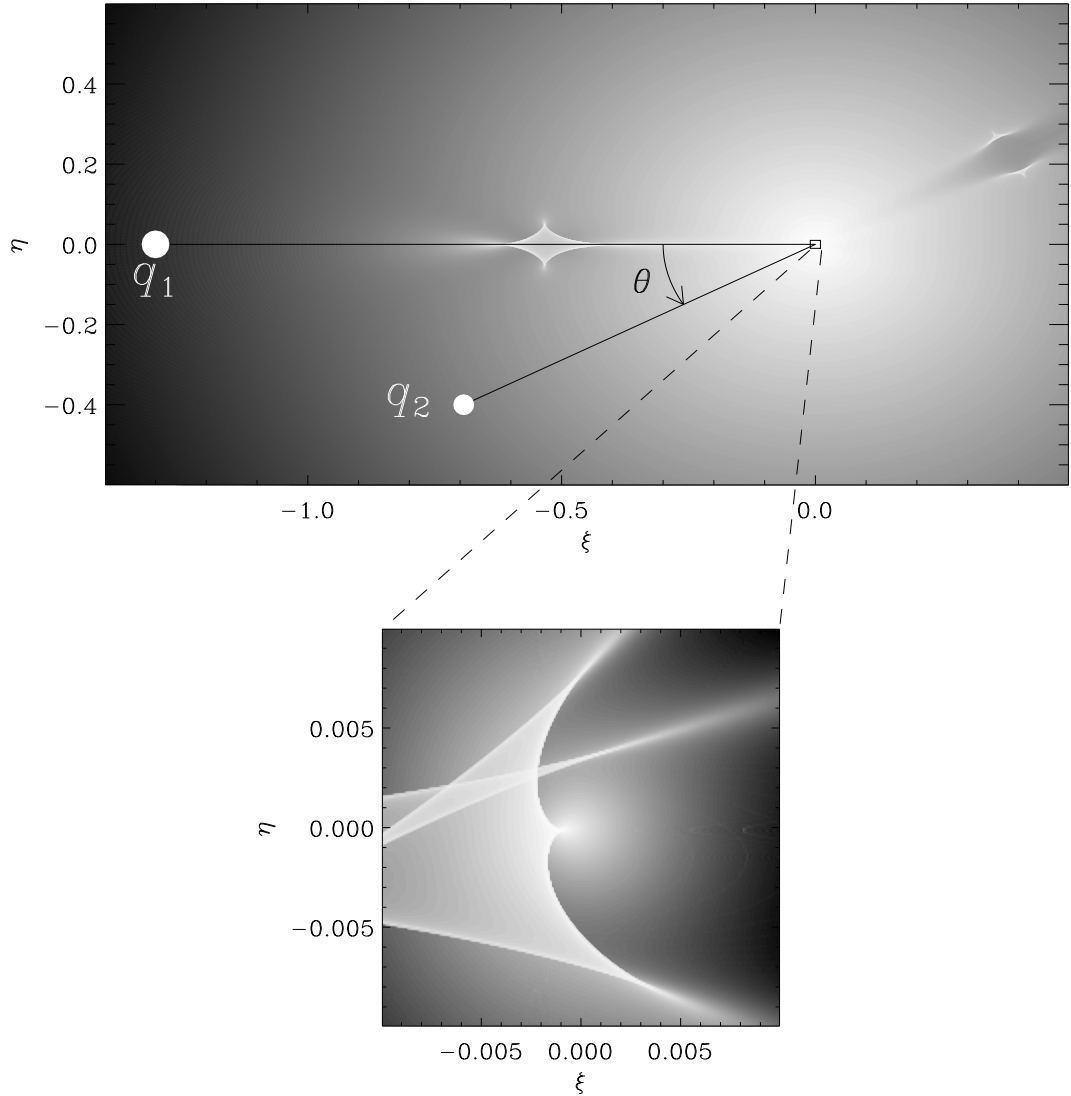}
\caption{Geometry of triple lens system composed of a lens star, a
primary planet and a secondary planet. In the upper panel, the
coordinate $(\xi,\eta)$ are centered at the position of the lens
star, and $\theta$ is the angle between position vectors of two
planets $(q_1,q_2)$. All lengths are normalized by the radius of the
Einstein ring. The gray scale represents the magnification and
brighter tone represents higher magnification. The lower panel shows
the zoom of the boxed region of the high magnification within
$u_0\leq0.01$.}
\end{center}
\end{minipage}
\end{figure*}

\begin{figure*}
\begin{minipage}[hpt]{17cm}
\begin{center}
\includegraphics[scale=1.0,angle=0,clip=true]{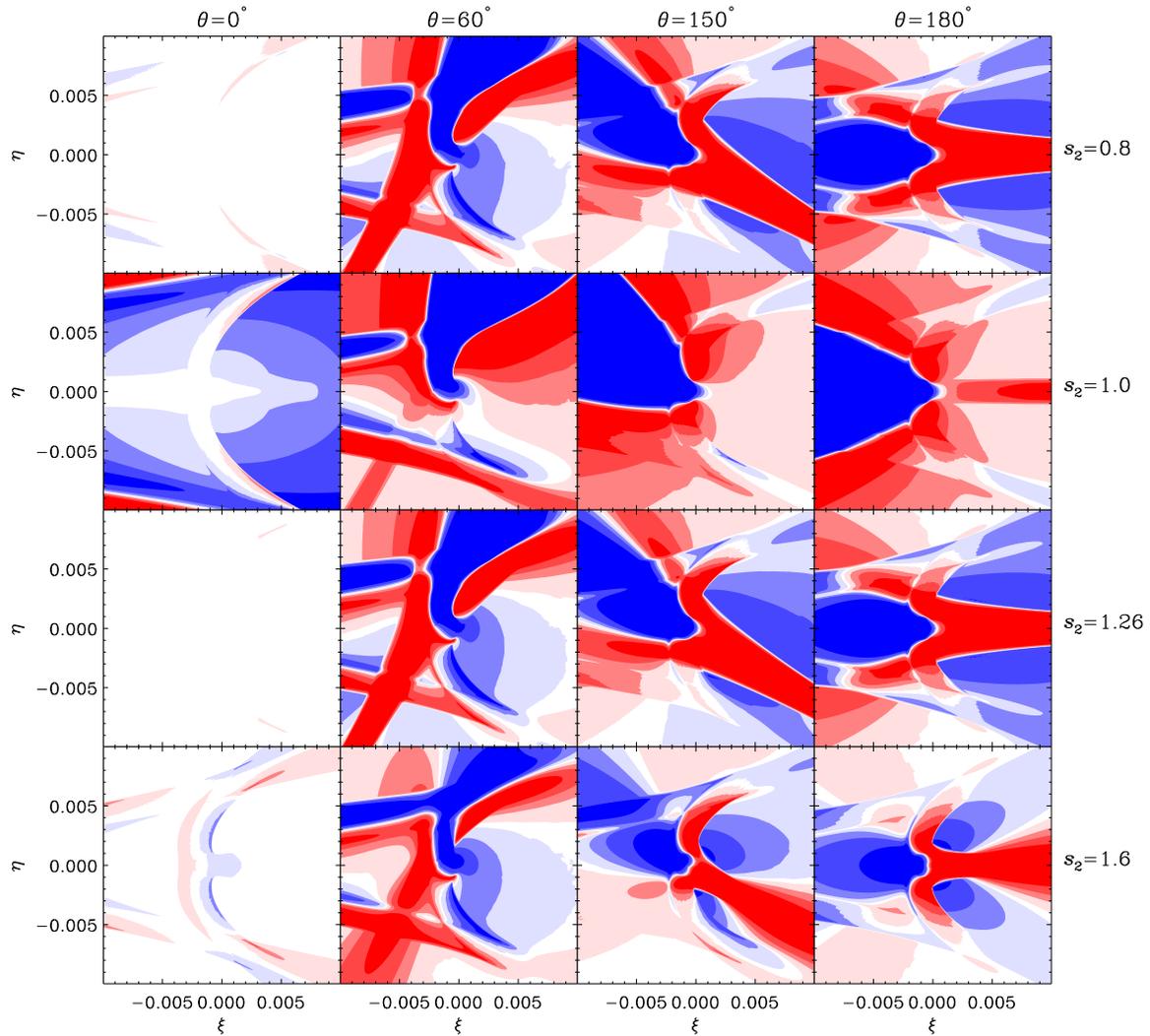}
\caption{Fractional deviation in magnification of triple lens system
    from the binary lens system. Deviation maps are presented
    in terms of the lens position of the secondary planet both in angles $\theta$ between position
    vectors of two planets and in separations $s_2$, each of which is calculated
    as a function of the source position $(\xi,\eta)$. The separation of the low-mass planet
    and the angle between two planets are marked on the right side of each row and the top of
    each column, respectively. The coordinates are set so that the lens star is at the center.
    The primary planet is fixed to be located on the $\xi$ axis, $(s_{1x},s_{1y})=(-1.3,0.0)$.
    We set the mass ratio of the primary planet to the lens star $q_1=3\times10^{-3}$
    and that of the secondary planet $q_2=1\times10^{-3}$.   Contours are drawn at the levels of
    $\varepsilon=\pm 1\%$, $\pm5\%$, $\pm10\%$, and $\pm20\%$, and the regions of negative
    $\varepsilon$ are shown in blue and positive $\varepsilon$ in red. The color changes into light shade as
    $|\varepsilon|$ level decreases.}
\end{center}
\end{minipage}
\end{figure*}

\begin{figure*}
\begin{minipage}[hpt]{17cm}
\begin{center}
\includegraphics[scale=1.0,angle=0,clip=true]{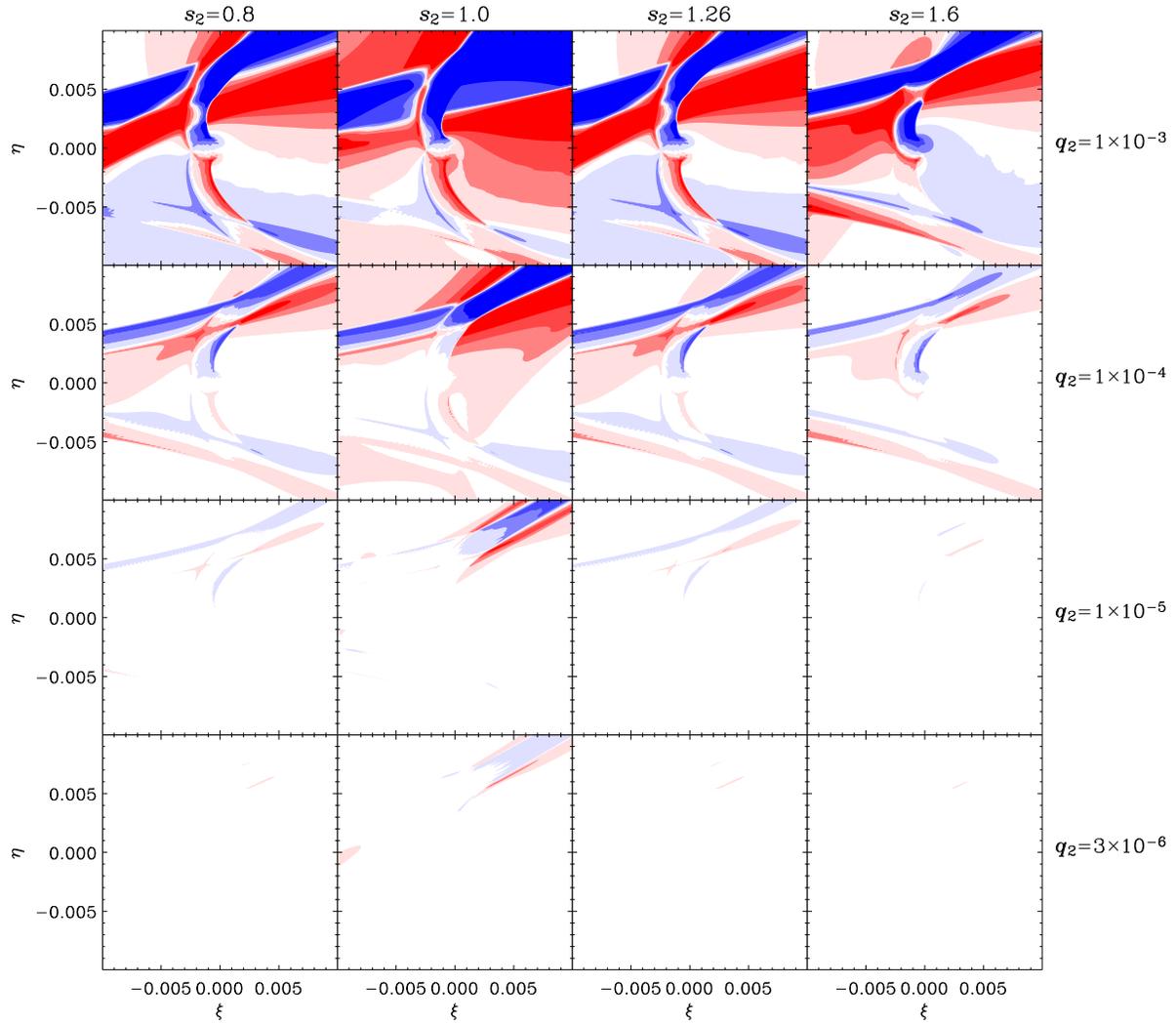}
\caption{Similar maps as Fig.~2, except that we set the angle
between position vectors of two planets  $\theta=30\degr$ and vary
the mass of the secondary planet. As in Fig.~2, contours are drawn
at the levels of $\varepsilon=\pm 1\%$, $\pm5\%$, $\pm10\%$, and
$\pm20\%$, and the regions of negative $\varepsilon$ are shown in
blue and positive $\varepsilon$ in red.}
\end{center}
\end{minipage}
\end{figure*}

\begin{figure*}
\begin{minipage}[hpt]{17cm}
\begin{center}
\includegraphics[scale=1.0,angle=0,clip=true]{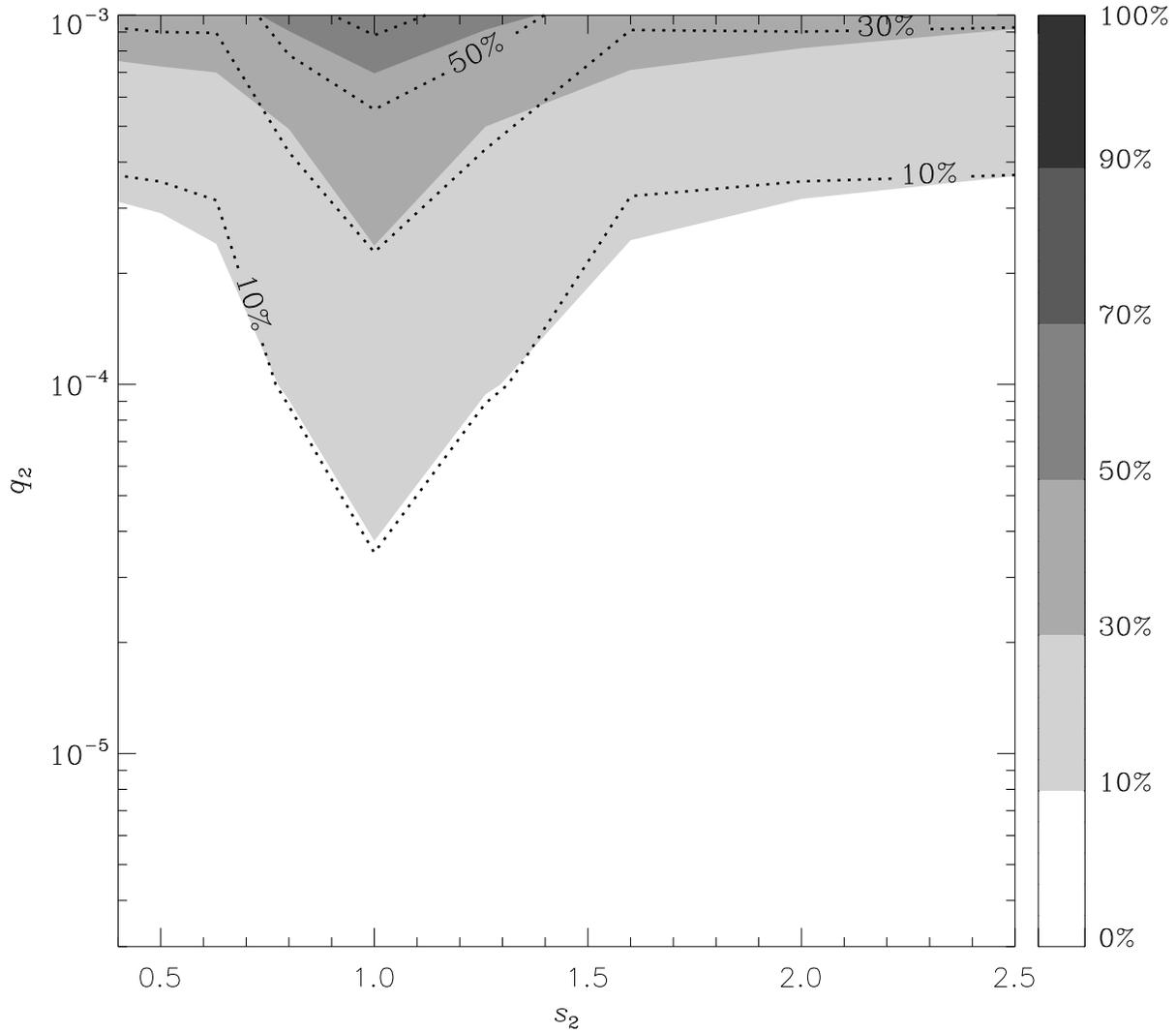}
\caption{Probabilities of detecting the low-mass planet in binary
systems and in triple systems. The detection probability of the
secondary low-mass planet in the triple lens system is represented
by grey-scale, and is drawn such that darker shade represents higher
probability as indicated in the gray index.    For comparison, we
also present as dotted contours the detection probability of
detecting the same low-mass planet if it is in a binary systems.
Probabilities are calculated such that the value of $|\varepsilon|$
in the deviation map  is greater than $5 \%$, considering only
$|u_0| \leq 0.01$ events.}
\end{center}
\end{minipage}
\end{figure*}

\begin{figure*}
\begin{minipage}[hpt]{17cm}
\begin{center}
\includegraphics[scale=1.0,angle=0,clip=true]{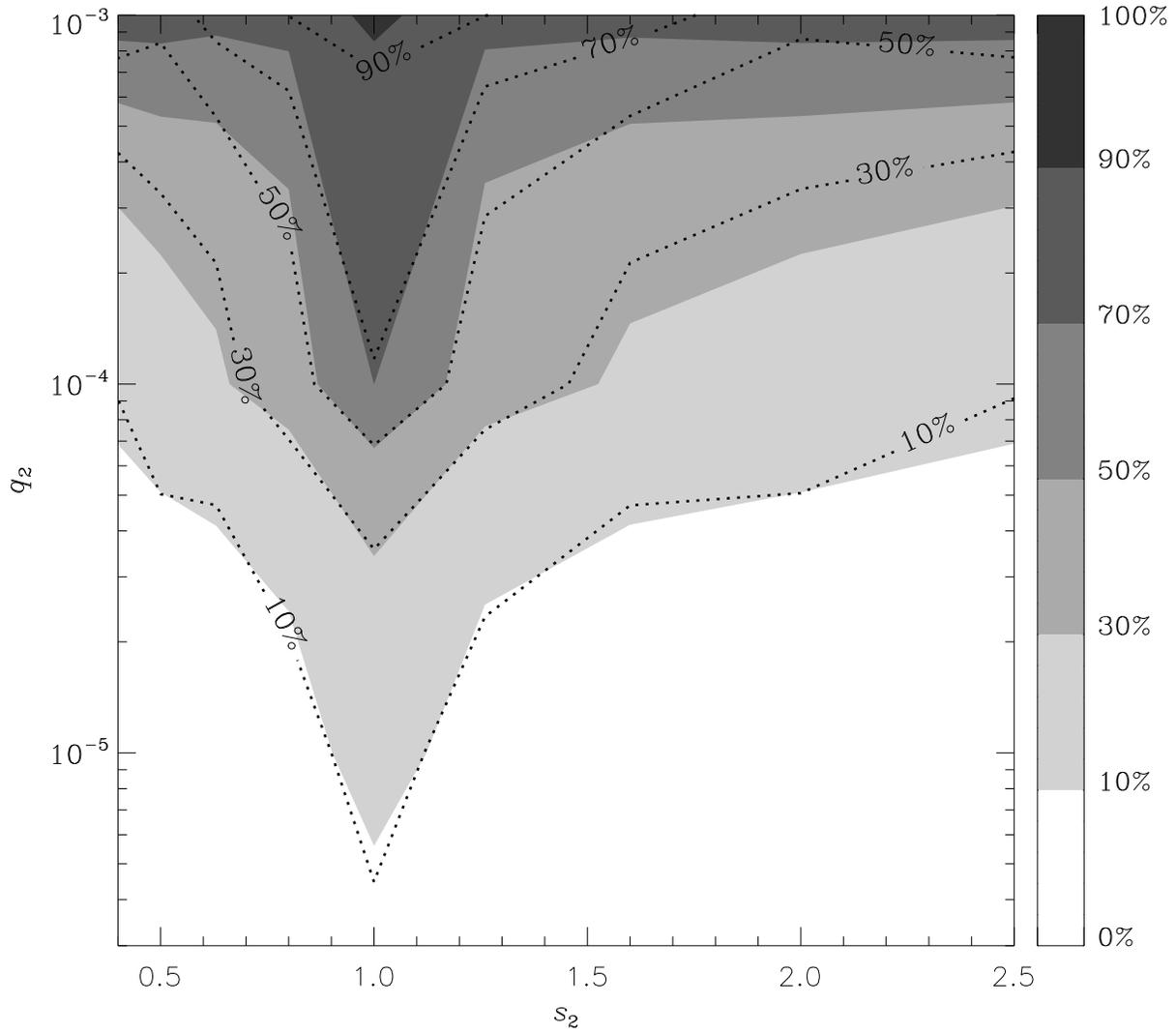}
\caption{Similar maps as Fig.~4, except that probabilities are
calculated such that the value of $|\varepsilon|$ in the deviation
map  is greater than $1 \%$, considering only $|u_0| \leq 0.01$
cases.}
\end{center}
\end{minipage}
\end{figure*}

\label{lastpage}

\end{document}